\documentclass[12pt]{article}

\usepackage{graphicx,color}
\usepackage{amsmath,amssymb,bm}
\usepackage{dcolumn}

\def\XXint#1#2#3{{\setbox0=\hbox{$#1{#2#3}{\int}$}
     \vcenter{\hbox{$#2#3$}}\kern-.5\wd0}}

\def\1{\'{\i}}

\begin{document}

\title{\bf The D model for deaths by COVID-19}

\author{J.E. Amaro \\ Departamento de
  F\'{\i}sica At\'omica, Molecular y Nuclear \\ and Instituto Carlos I
  de F{\'\i}sica Te\'orica y Computacional \\ Universidad de Granada,
  E-18071 Granada, Spain.\\
amaro@ugr.es }

\maketitle

\begin{abstract}
  \rule{0ex}{3ex}

We present a simple analytical model to describe the fast increase of
deaths produced by the corona virus (COVID-19) infections. The
'D' (deaths)  model comes from a simplified version of the SIR
(susceptible-infected-recovered) model known as SI model.
It assumes that there is no
recovery. In that case the dynamical equations can be solved
analytically and the result is extended to describe the
D-function that depends on three parameters
that we can fit to the data.  Results for the data from Spain, Italy
and China are presented.  The model is validated by comparing with the
data of deaths in China, which are well described. This allows to make
predictions for the development of the disease in Spain and Italy.

\end{abstract}

\section{Introduction}

The SIR (susceptible-infected-recovered) model, developed by Ross,
Hamer, and others \cite{And91}, is widely used among the
many epidemiological models as a first approach to virus spreading, with
applications to many other sociological situations \cite{Rod16}.  It
consists of a system of three coupled non-linear ordinary differential
equations \cite{Wei13} involving three time-dependent functions:
\begin{itemize}
  \item Infected individuals, $I(t)$.
  \item Susceptible individuals, $S(t)$.
  \item Recovered individuals, $R(t)$.
\end {itemize}
The resulting dynamical system is the following
\begin{eqnarray}
  \frac{dS}{dt} & = & - \lambda S I \\
  \frac{dI}{dt} & = &  \lambda S I -\beta I \\
  \frac{dR}{dt} & = & \beta I \\
  S(t)+I(t)+ R(t) &=& N
 \end{eqnarray}
where $\lambda >0$ is the corona virus transmission rate, $\beta >0$ is the
recovery rate, and $N$ is the total population size.  The system is
reduced to two coupled differential equations, which does not possess
an explicit formula solution, but can be solved numerically. The SIR
model is usually parametrized using actual infection data and the
solution of the $I(t)$ function can be compared with actual infection
data, to predict the evolution of the disease.

In this paper we make drastic assumptions in order to obtain an
analytical formula to describe the evolution of deaths by corona virus.
This can be useful as a fast method to foresee the global behavior as
a first approach before applying more sophisticated methods.  We shall
see that the resulting 'D' (deaths) model, that derivates from the SI model (no recovery), and is extended to deaths, describes well enough the
data of the current corona virus pandemic in the countries China, Spain
and Italy, where the pandemic is stronger.

\section{The D model}

The first basic assumption of the model is that there is no recovery from
corona virus, at least during the pandemic time interval. This drastic
assumption could be reasonable if the spreading time of the pandemic
is much faster than the recovery time, or  $\beta \ll \lambda$. 

Under this simple assumption $R(t)=0$, and the SIR equations reduce to
the  single equation of the well known SI model
\begin{equation}
  \frac{dI}{dt}  =   \lambda (N-I(t)) I(t).  \\
\end{equation}
Therefore the infection rate is proportional to the infected, $I$
and to the non-infected $N-I$ or susceptible individuals.

This equation is easily solved in the following way. Dividing by $(N-I)I$
and multiplying by $dt$,
\begin{equation}
\frac{dI}{(N-I)I} = \lambda dt
\end{equation}
or
\begin{equation}
  \left(
  \frac{1}{N-I}
+  \frac{1}{I}
\right)dI
= \lambda N dt
\end{equation}  
Integrating over $t=0$ and $t$ we obtain
\begin{equation}
  \ln\frac{I}{N-I} - 
  \ln\frac{I_0}{N-I_0} = \lambda N (t-t_0)
\end{equation}
Where $I_0=I(t_0)$. Taking the exponential on both sides
\begin{equation}
  \frac{I}{N-I} =   \frac{I_0}{N-I_0} {\rm e}^{\lambda N (t-t_0)}
\end{equation}
Finally, solving this algebraic equation we obtain the solution $I(t)$
\begin{equation}
  I(t) = \frac{ N I_0  {\rm e}^{\lambda N (t-t_0)}  }
  {  N-I_0 +  I_0  {\rm e}^{\lambda N (t-t_0)}  }
\end{equation}
We write this equation in the form
\begin{equation}
  I(t) =  \frac{ I_0 \,{\rm e}^{ (t-t_0)/b}  }
  {  1-C +  C \, {\rm e}^{(t-t_0)/b}  }
\end{equation}
Where we have defined the constants
\begin{equation}  \label{parametros}
  b= \frac{1}{\lambda N}, \kern 1cm C = \frac{I_0}{N}
  \end{equation}
The parameter $b$ is the characteristic evolution time of the initial
exponential increase of the pandemic. The constant $C$ is the initial
infestation rate (with respect to the total population). Assuming that initially $C \ll 1$, this constant can be neglected in the denominator, obtaining
\begin{equation}
  I(t) =  \frac{ I_0 \,{\rm e}^{ (t-t_0)/b}  }
  {  1+  C \, {\rm e}^{(t-t_0)/b}  }
\end{equation}

Now to predict the number of deaths in the D model we assume that the
number of deaths at some time $t$ is proportional to the infestation
at some former time $\tau$, that is,
\begin{equation}
  D(t) = m I(t-\tau)
\end{equation}
Where $m$ is the mortality  or death rate, and $\tau$ is the mortality time.

With this assumption we can finally write the D model equation as
\begin{equation} \label{D-model}
  D(t) =  \frac{ a {\rm e}^{ (t-t_0)/b}  }
  {  1 +  c \, {\rm e}^{(t-t_0)/b}  }
\end{equation}
where $a= m I_0 \, {\rm e}^{-\tau/b} $, and  
 $c = C \, {\rm e}^{-\tau/b} $.  
This is the final equation for the model. This simple function has three
parameters, $a,b,c$, which we fit to the data.
Note that the rest of the parameters, $m$, $\tau$, $I_0$ and $N$
are not needed in our model because
they are included inside the fitted parameters $a,b,c$.

\begin{figure}
\begin{center}
\includegraphics[width= 8cm, bb=170 550 430 780]{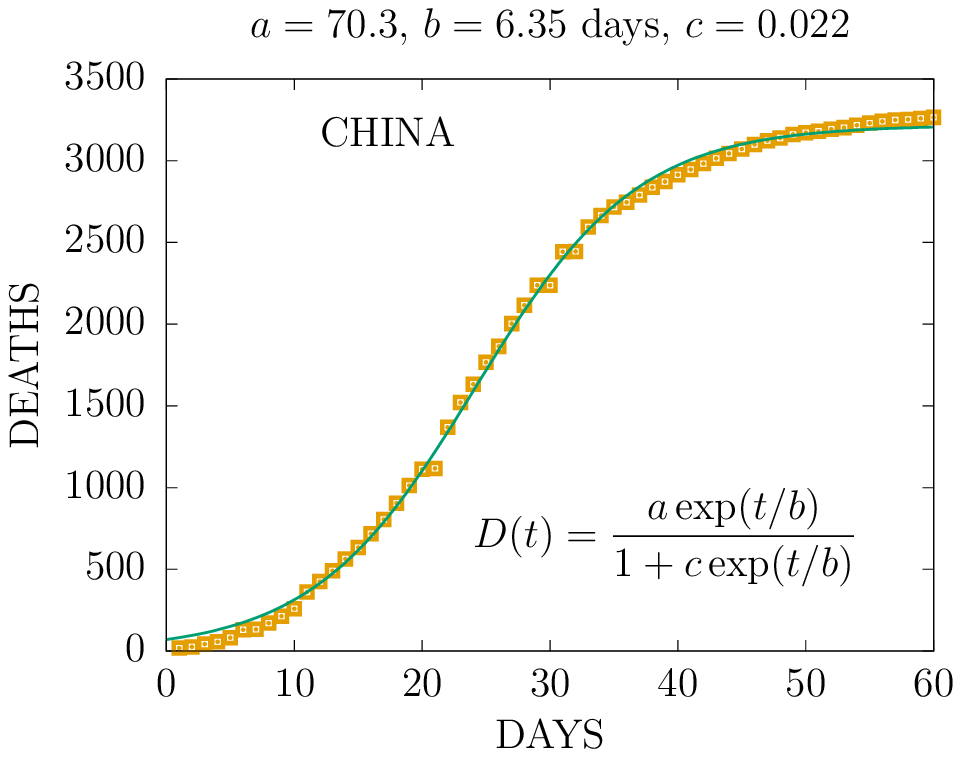}
\caption{Fit of the D-model to the deaths by corona virus in China.
  Data are from \cite{worldometer}.
}
\end{center}
\end{figure}

\section{Results}

In this section we fit the three parameters, $a,b,c$ of our model,
Eq. (\ref{D-model}), to real data for three countries, China, Spain
and Italy. The data are taken from ref. \cite{worldometer}

In Fig, 1 we show our fit of the D-model to the death data of
China. In this country the evolution has been apparently controlled and
the D function has already arrived to the plateau zone, with few
increments over time, or fluctuations that are beyond the model
assumptions. We see that the pandemic lasted for about two months to
reach the top end of the curve. Fig. 1 shows that the model describes
well the COVID-19 evolution of deaths, despite our crude assumptions.
This validation allows us to trust its applicability in other cases
where the pandemic is still in its initial phase, to make predictions.

\begin{figure}
\begin{center}
\includegraphics[width= 8cm, bb=170 550 430 780]{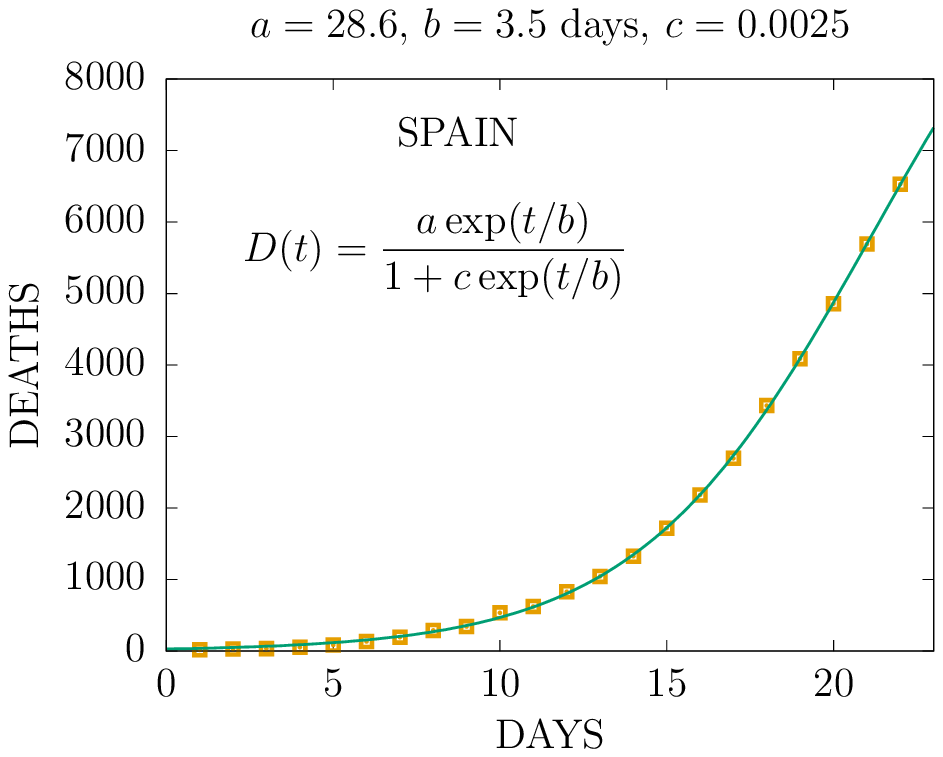}
\caption{Fit of the D-model to the deaths by corona virus in Spain.
  Data are from \cite{worldometer}.
}
\end{center}
\end{figure}

\begin{figure}
\begin{center}
\includegraphics[width= 8cm, bb=170 550 430 780]{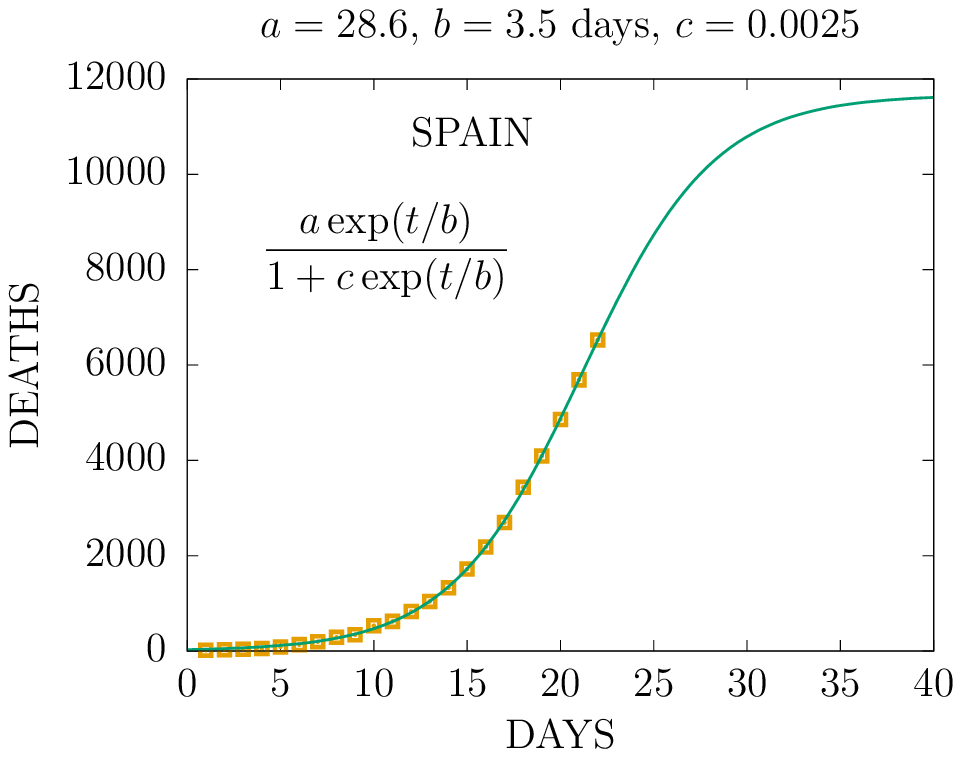}
\caption{Prediction of the D-model to the COVID-19 pandemic in Spain.
  Data are from \cite{worldometer}.
}
\end{center}
\end{figure}

\begin{figure}
\begin{center}
\includegraphics[width= 8cm, bb=170 550 430 780]{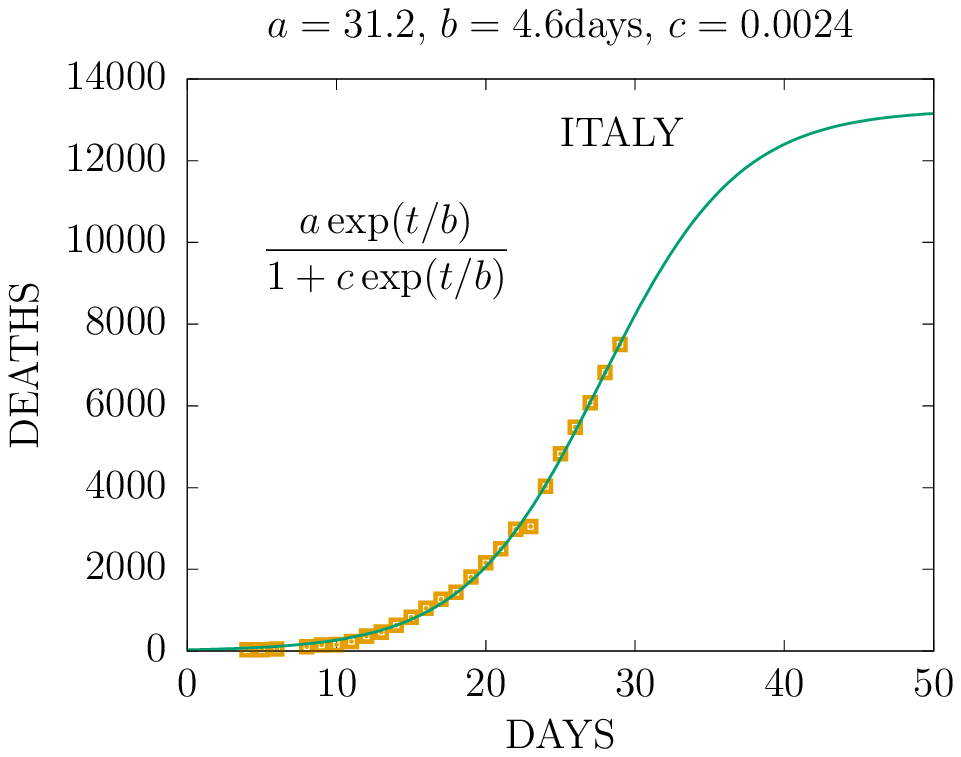}
\caption{Prediction of the D-model to the COVID-19 pandemic in Italy.
  Data are from \cite{worldometer}.
}
\end{center}
\end{figure}

In Fig, 2 we show our fit of the D-model to the death data of
Spain. The data start on March, 8 2020 (day 1).  In this country the
evolution has been strong and up to day 22 (March 29 2020) they have
exceeded the deaths in china, reaching almost 7000 deaths.  In Fig. 2
we see that the D-model describes quite well this evolution region up
to day 22 of the pandemic.

In fig.3 we show our prediction for the next weeks in Spain, using the
fit to the first 22 days. According to the figure we reached
the mid
point of the curve on day 21.
It is expected that Spain will reach the top
plateau of the curve in 18 days, that is in mid April.

Finally in Fig. 4 we show our fit and prediction to the Italy data.
The parameters are similar to the case of Spain with about 20 days
to the top of the curve.

\section{Final remarks}

To conclude we have seen that the D-model for COVID-19 pandemic,
derived form the SI model, describes well the current data of China,
Spain and Italy with only three parameters.

The assumption made here, that the recovered individual do not
influence the increase of the infected, could further indicate that
the total
population $N$ is not a constant as assumed in the SIR model, but it
increases over time as more people are exposed, for example, in
villages that until now had been isolated from the sources of
infection in big cities.

The D model is simple enough to provide 
fast estimations of pandemic evolution in other countries, and could
be useful for the control of the disease.

\section{Acknowledgements}

The author thanks useful comments from Nico Orce and from the
WhatsApp group Covid-19.

This work is supported by Spanish Ministerio de Economia y
Competitividad and European FEDER funds (grant FIS2017-85053-C2-1-P)
and Junta de Andalucia (grant FQM-225).


\end{document}